\documentclass[10pt]{article}
\usepackage{fullpage}
\RequirePackage{algorithm,algorithmic,amssymb,epsf,epic,amsmath}

\newtheorem{theorem}{Theorem}[section]

\newtheorem{lemma}[theorem]{Lemma}

\newtheorem{comment}[theorem]{Comment}

\def\boldhead#1:{\par\vskip 7pt\noindent{\bf #1:}\hskip 10pt}
\def\ithead#1:{\par\vskip 7pt\noindent{\it #1:}\hskip 10pt}

\def\inline#1:{\par\vskip 7pt\noindent{\bf #1:}\hskip 10pt}
\def\midinline#1:{\par\noindent{\bf #1:}\hskip 10pt}
\def\dnsinline#1:{\par\vskip -7pt\noindent{\bf #1:}\hskip 10pt}
\def\ddnsinline#1:{\newline{\bf #1:}\hskip 10pt}
\def\largeinline#1:{\par\vskip 7pt\noindent{\large\bf #1:}\hskip 10pt}
\long\def\comment #1\commentend{}
\long\def\commhide #1\commhideend{}
\long\def\commfull #1\commend{#1}
\long\def\commabs #1\commenda{}
\long\def\commtim #1\commendt{#1}
\long\def\commb #1\commbend{}
\long\def\commedit #1\commeditend{} % Editing comments, marked also by $>>>$ 

\long\def\commB #1\commBend{}       % Omit in 1996 (both TR and Siena)
\long\def\commex #1\commexend{}     % LN home exercise (hide solutions)

\long\def\commsiena #1\commsienaend{}  % omit in Siena, show in TR
                                         
\long\def\commBI #1\commBIend{}  % omit in Bar-Ilan
                                         
\long\def\CProof #1\CQED{}

\def\blackslug{\hbox{\hskip 1pt \vrule width 4pt height 8pt
    depth 1.5pt \hskip 1pt}}
\def\QED{\quad\blackslug\lower 8.5pt\null\par}
\def\Proof{\par\noindent{\bf Proof:~}}

\def\proof{\Proof}

\long\def\PPP#1{\noindent{\bf Proof:}{ #1}{\quad\blackslug\lower 8.5pt\null}}

\long\def\denspar #1\densend
{#1}
\newif\ifnotesw\noteswtrue% T to show box & marginal notes; F supresses.
   {\ifnotesw\marginpar[\hfill\(\top\)]{\(\top\)}\fi}%
   {\ifnotesw\marginpar[\hfill\(\bot\)]{\(\bot\)}\fi}

\newcommand{\mnote}[1]%
    {\ifnotesw\marginpar%
        [{\scriptsize\it\begin{minipage}[t]{\marginparwidth}
        \raggedleft#1%
                        \end{minipage}}]%
        {\scriptsize\it\begin{minipage}[t]{\marginparwidth}
        \raggedright#1%
                        \end{minipage}}%
    \fi}

\def\cB{{\cal B}}

\def\cE{{\cal E}}

\def\cG{{\cal G}}

\def\cP{{\cal P}}

\def\cU{{\cal U}}

\def\hC{{\hat C}}

\def\hQ{{\hat Q}}

\def\tG{{\tilde G}}

\def\tO{{\tilde O}}

\def\MathF{\hbox{\rm I\kern-2pt F}}
\def\MathP{\hbox{\rm I\kern-2pt P}}
\def\MathR{\hbox{\rm I\kern-2pt R}}
\def\MathZ{\hbox{\sf Z\kern-4pt Z}}
\def\MathN{\hbox{\rm I\kern-2pt I\kern-3.1pt N}}
\def\MathC{\hbox{\rm \kern0.7pt\raise0.8pt\hbox{\footnotesize I}
\kern-4.2pt C}}
\def\MathQ{\hbox{\rm I\kern-6pt Q}}

\newsavebox{\ttop}\newsavebox{\bbot}

\def\eps{\epsilon}

\def\setmns{\setminus}

\def\nin{{~\not \in~}}
\def\emset{\emptyset}

\def\etal{\emph{et~al.}}

\usepackage{multirow,graphicx}
\usepackage{boldline}

\usepackage{pdfsync,fullpage,comment}
\usepackage{authblk}

\ifx\pdftexversion\undefined
\usepackage[colorlinks,linkcolor=black,filecolor=black,citecolor=black,urlco
lor=black,pdfstartview=FitH]{hyperref}
\else
\usepackage[colorlinks,linkcolor=blue,filecolor=blue,citecolor=blue,urlcolor
=blue,pdfstartview=FitH]{hyperref}
\fi

\newcommand{\alert}[1]{\textbf{\color{red}
[[[#1]]]}\marginpar{\textbf{\color{red}**}}\typeout{ALERT:
\the\inputlineno: #1}}

\newcommand{\namedref}[2]{\hyperref[#2]{#1~\ref*{#2}}}

\newcommand{\R}{\mathbb{R}}
\newcommand{\E}{{\mathbb{E}}}
\begin{document}
\def\hpi{\hat{\pi}}
\def\rt{\mathit{rt}}
\def\hd{\hat{d}}
\def\chS{\hat{\cal S}}
\def\chP{\hat{\cal P}}
\def\chU{\hat{\cal U}}
\def\exp{\mathit{exp}}
\def\Bunch{\mathit{Bunch}}
\def\Rt{\mathit{Roots}}
\def\Lab{\mathit{Label}}
\def\Id{\mathit{Id}}
\def\Lev{\mathit{Level}}
\def\URt{\mathit{URoots}}
\def\Ball{\mathit{Ball}}
\def\wmax{{w_{max}}}
\def\tO{\tilde{O}}
\def\nin{\not \in}
\def\emset{\emptyset}
\def\setmns{\setminus}
\def\etal{{et al.~}}
\def\Pairs{\mathit{Pairs}}
\def\Paths{\mathit{Paths}}
\def\pred{\mathit{pred}}
\def\Rad{\mathit{Rad}}
\def\succ{\mathit{succ}}
\def\NULL{\mathit{NULL}}
\def\exp{\mathit{exp}}
\def\Ball{\mathit{Ball}}
\def\tPi{\tilde{\Pi}}
\def\deg{\mathit{deg}}
\def\TB{\cB^{{1/3}}}
\def\Branch{\mathit{Branch}}
\def\uzero{u^{(0)}}
\def\uone{u^{(1)}}
\def\uj{u^{(j)}}
\def\vzero{v^{(0)}}
\def\vone{v^{(1)}}
\def\vj{v^{(j)}}
\def\dzero{d^{(0)}}
\def\done{d^{(1)}}
\def\dj{d^{(j)}}
\def\dG{d_G}
\def\third{{1 \over 3}}
\def\stretchexp{{\log_{4/3} 7}}
\def\tLambda{\tilde{\Lambda}}
\def\tomega{\tilde{\omega}}
\def\HKNfactor{ 2^{\tO(\sqrt{\log \max \{n,\Lambda\}})}}
\def\ourfactor{ (1/\eps)^{O(\sqrt{{\log n} \over {\log\log n}})} \cdot 2^{O(\sqrt{\log n \cdot \log\log n})} }
\def\hQ{{\hat{Q}}}

\newcommand{\Patrascu}{P\v{a}tra\c{s}cu{~}}
\newcommand{\Lists}{{\rm Lists}}
\newcommand{\td}{{\tilde{d}}}
\title{Near-Additive Spanners and Near-Exact Hopsets, A Unified View}

\author[1]{Michael Elkin\thanks{This research was supported by the ISF grant No. (724/15).}}
\author[1]{Ofer Neiman\thanks{Supported in part by ISF grant No. (1817/17) and by BSF grant No. 2015813.}}
\affil[1]{Ben-Gurion University of the Negev. Email: \texttt{\{elkinm,neimano\}@cs.bgu.ac.il}}

\date{}
\maketitle

\begin{abstract}
Given an {\em unweighted} undirected graph $G = (V,E)$, and a pair of parameters $\eps > 0$, $\beta = 1,2,\ldots$,
a subgraph $G' =(V,H)$, $H \subseteq E$, of $G$ is a {\em $(1+\eps,\beta)$-spanner} (aka, a {\em near-additive spanner}) of $G$ if for every $u,v \in V$,
$$d_{G'}(u,v) \le (1+\eps)d_G(u,v) + \beta~.$$
It was shown in \cite{EP01} that for any $n$-vertex $G$ as above, and any $\eps > 0$ and $\kappa = 1,2,\ldots$, there exists a $(1+\eps,\beta)$-spanner $G'$ with $O_{\eps,\kappa}(n^{1+1/\kappa})$ edges, with
$$\beta = \beta_{EP} = \left({{\log \kappa} \over \eps}\right)^{\log \kappa - 2}~.$$
This bound remains state-of-the-art, and its dependence on $\eps$ (for the case of small $\kappa$) was shown to be tight in \cite{ABP18}.

Given a {\em weighted} undirected graph $G = (V,E,\omega)$, and a pair of parameters $\eps > 0$, $\beta = 1,2,\ldots$, a graph $G'= (V,H,\omega')$ is a {\em $(1+\eps,\beta)$-hopset} (aka, a  {\em near-exact hopset}) of $G$ if for every $u,v \in V$,
$$d_G(u,v) \le d_{G\cup G'}^{(\beta)}(u,v) \le (1+\eps)d_G(u,v)~,$$
where $ d_{G\cup G'}^{(\beta)}(u,v)$ stands for a $\beta$-(hop)-bounded distance between $u$ and $v$ in the union graph $G \cup G'$.
It was shown in \cite{EN16} that for any $n$-vertex $G$ and $\eps$ and $\kappa$ as above, there exists a
$(1+\eps,\beta)$-hopset with $\tO(n^{1+1/\kappa})$ edges, with $\beta = \beta_{EP}$.

Not only the two results of \cite{EP01} and \cite{EN16} are strikingly similar, but so are also their proof techniques.
Moreover, Thorup-Zwick's later construction of near-additive spanners \cite{TZ06} was also shown in \cite{EN19,HP17} to provide hopsets with analogous (to that of \cite{TZ06}) properties.

In this survey we explore this intriguing phenomenon, sketch the basic proof techniques used for these results, and highlight open questions.
\end{abstract}

%\end{titlepage}

\thispagestyle{empty}
\newpage
\setcounter{page}{1}
%\pagenumbering {arabic} % back to usual

%%%%%%%%%%%%%%%%%%%%%%%%
\section{Introduction}
%%%%%%%%%%%%%%%%%%%%%%%%

%%%%%%%%%%%%%%%%%%%%%%%%%%
\subsection{Spanners}
\label{sec:intro_sp}
%%%%%%%%%%%%%%%%%%%%%%%%%%%

Given an undirected unweighted $n$-vertex graph $G = (V,E)$, and a pair of paremeters $\alpha \ge 1, \beta \ge 0$, a subgraph $G'= (V,H)$, $H \subseteq E$, of $G$ is called an {\em $(\alpha,\beta)$-spanner} of $G$, if for every pair $u,v \in V$ of vertices, we have $d_H(u,v) \le \alpha \cdot d_G(u,v) + \beta$.  Here $d_G$ (respectively, $d_H$) stands for the distance in $G$ (respectively, in $H$). If $\beta = 0$, the spanner is called {\em multiplicative}, and if $\alpha=1$, the spanner is called {\em additive}.  A graph $G' = (V,H,\omega)$ (possibly weighted, even when the original graph $G = (V,E)$ is unweighted) is called an {\em $(\alpha,\beta)$-emulator} of $G$, if for every pair $u,v \in V$, we have
$d_G(u,w) \le d_H(u,v) \le \alpha d_G(u,v) + \beta$.

Althofer et al. \cite{ADDJ90}, improving upon an earlier work by Peleg and Schaeffer \cite{PS89}, showed that for every $n$-vertex undirected (possibly weighted) graph $G = (V,E)$, and any parameter $\kappa = 1,2,\ldots$, there exists a $(2\kappa-1)$-spanner with at most $n^{1+1/\kappa}$ edges. This bound is known to be tight under Erdos' girth conjecture
(see, e.g., \cite{TZ01}, Section 5), and is unconditionally tight up to a leading constant coefficient in the stretch.

A large body of literature exploring constructions of spanners in various computational settings was developed throughout the years \cite{ABCP93,Cohen98,ACIM99,HZ96-c,DHZ00,EP01,E01,BS03,EZ04,TZ06,Pet07,P08,P10,Wood06,BKMP05,CW04,RTZ05,C13,AB16,
ABP18,EN17,EM19}.
Algorithms for constructing purely additive spanners were given in \cite{ACIM99,DHZ00,EP01,BKMP05,C13}. Specifically, Aingworth et al. and Dor et al. \cite{ACIM99,DHZ00} devised an algorithm constructing additive 2-spanners with $\tO(n^{3/2})$ edges, and additive 4-emulators with $\tO(n^{4/3})$ edges. Elkin and Peleg \cite{EP01} shaved polylogarithmic factors from these size bounds via different constructions; their size bounds are $O(n^{3/2})$ and $O(n^{4/3})$, respectively. Baswana et al. \cite{BKMP05} devised a construction of additive 6-spanners with $O(n^{4/3})$ edges, and Chechik \cite{C13} devised a construction of additive 4-spanners with $\tO(n^{7/5})$ edges.

In \cite{EP01} Elkin and Peleg also devised the first construction of {\em near-additive} spanners, i.e., $(1+\eps,\beta)$-spanners, which are together with near-exact hopsets, constitute the main topic of the current survey. Specifically, they showed that for any $\eps > 0$ and $\kappa = 1,2,\ldots$, there exists $\beta(\eps,\kappa) $ (denoted also $\beta_{EP}$) such that for any $n$-vertex unweighted undirected graph there exists a $(1+\eps,\beta)$-spanner with $O_{\eps,\kappa}(n^{1+1/\kappa})$ edges.

Note that in this result, unlike in the aforementioned tradeoff for multiplicative spanners \cite{PS89,ADDJ90}, both the multiplicative stretch $1+\eps$ and the exponent of the number of edges $1 + 1/\kappa$ can {\em simultaneously} be made as small as one wishes, at the expense of increasing the additive error term $\beta$. This additive term behaves as
$$\beta_{EP}(\eps,\kappa) = \left({{\log \kappa} \over \eps}\right)^{\log \kappa - 2}~,$$
and it is still the state-of-the-art bound.

Observe also that this result means also that distances larger than some constant threshold can be approximated arbitrarily well using arbitrarily sparse spanners. The threshold increases, of course, as the approximation factor and the exponent of the number of edges decrease.

At the beginning near-additive spanners were often viewed as a stepping stone towards the ``real thing", that is, purely additive spanners. However, Abboud  and Bodwin \cite{AB16}, relying on earlier lower bounds for distance preservers \cite{CE05}, showed that one cannot in general have purely additive spanners with constant (or even polylogarithmic) error term $\beta$ and size $o(n^{4/3})$. Therefore, near-additive spanners is the {\em best one can hope for}!

Near-additive spanners were intensively studied in the last two decades \cite{EP01,E01,EZ04,TZ06,Pet07,P08,P10,EN17,ABP18,EM19}.
In \cite{E01} Elkin devised the first efficient algorithm for constructing them.  This algorithm provides $(1+\eps,\beta)$-spanners with $\tO_{\eps,\kappa,\rho}(n^{1+1/\kappa})$ edges in centralized time $O(|E| \cdot n^\rho)$, with $\beta_E = (\kappa/\eps)^{O({{\log \kappa} \over \rho})}$, where $\rho > 0$ is an additional parameter that controls the running time.
Improved variants of this algorithm, as well as efficient implementations of it in distributed and streaming settings, were devised in \cite{EZ04}. Both these algorithms \cite{E01,EZ04} build upon ideas from a seminal algorithm of Cohen \cite{C94} for constructing hopsets.
(See more about it in Section \ref{sec:intro_hop}.)

Another remarkable algorithm for constructing near-additive spanners and emulators was devised by Thorup and Zwick \cite{TZ06}. The main feature of their construction is that it provides a {\em universal} near-additive spanner, i.e., the same spanner applies {\em simultaneously} for all values of $\eps > 0$.
Putting it differently, their algorithm accepts as input an $n$-vertex graph $G = (V,E)$ and a parameter $\kappa = 2,3,\ldots$, (but it does {\em not} accept $\eps$ as a part of the input), and it constructs for it a spanner $G' = (V,H)$, $H \subseteq E$, $|H| = O_\kappa(n^{1+1/\kappa})$, which constitutes a $(1+\eps,\beta(\eps,\kappa))$-spanner for all $\eps >0$ simultaneously.  The additive term in their construction behaves as
$\beta = \beta_{TZ} = O\left({\kappa \over \eps}\right)^{\kappa-1}$, i.e., it is much higher than
$\beta_{EP}$. On the other hand, they have also devised a {\em universal emulator} whose additive term is the same as in the {\em spanner} of \cite{EP01}.

Interestingly, the universality of Thorup-Zwick's construction enables one to obtain spanners and emulators with a {\em sublinear} additive error term. For concreteness, let us consider a pair of vertices $u,v \in V$ with $d_G(u,v) = d$. We can set
$$\eps = {{\log \kappa} \over d^{{1 \over {\log \kappa - 1}}}}~.$$
Then we have
$$d_H(u,v) \le d(1+\eps) + \left({{\log \kappa} \over \eps}\right)^{\log \kappa-2} \le
d + O\left(\log \kappa \cdot d^{1 - {1 \over {\log \kappa - 1}}}\right)~.$$
Note that the additive error $O\left(\log \kappa \cdot d^{1 - {1 \over {\log \kappa - 1}}}\right)$ is sublinear in the original distance $d = d_G(u,v)$, and this property holds for all pair of vertices.

Pettie \cite{Pet07,P08,P10} improved the construction of universal spanners of \cite{TZ06}.
His algorithm constructs universal $(1+\eps,\beta)$-spanners with $O_\kappa(n^{1+1/\kappa})$ edges and
$$\beta = \beta_{Pet} = \beta_{EP}^{\log_{4/3} 2} \approx \beta_{EP}^{2.41}~.$$
Devising universal spanners with additive error that matches the additive error of spanners of \cite{EP01} (i.e., closing the gap between $\beta_{Pet}$ and $\beta_{EP}$) is an open problem.
The algorithm of Pettie \cite{Pet07,P08,P10} is based on a combination of Thorup-Zwick's construction of emulators with a construction of distance preservers from \cite{CE05}.

Finally, Abboud et al. \cite{ABP18} showed a lower bound on the size-stretch tradeoff of near-additive spanners. They showed that any construction of $(1+\eps,\beta)$-spanners with
$O(n^{1+1/\kappa})$ edges must have
$$\beta_{ABP} = \Omega\left({1 \over {\eps \cdot \log \kappa}}\right)^{\log \kappa-2}~.$$
Note that while this  lower bound is tight for a very small $\eps > 0$ and constant $\kappa$,
it is meaningless when $\eps \ge {1 \over {\log \kappa}}$. So, in particular, it is wide open if one can achieve near-additive spanners with $\beta$ that depends {\em polynomially} on $\kappa$.
(The state-of-the-art dependence is $\beta_{EP} = (\log \kappa)^{\log \kappa} = \kappa^{\log\log \kappa}$, i.e., it is slightly superpolynomial in $\kappa$.)

We note that if one allows a larger but still constant multiplicative stretch, then
$(O(1),\beta)$-spanners with $\tO_\kappa(n^{1+1/\kappa})$ edges with $\beta = \mathit{poly}(\kappa)$ were devised by Pettie \cite{Pet07}. Improved and generalized bounds along these lines were given in \cite{EGN19,BLP19}.

%%%%%%%%%%%%%%%%%%%%%%%%%
\subsection{Near-Exact Hopsets}
\label{sec:intro_hop}
%%%%%%%%%%%%%%%%%%%%%%%%%%%%

Given an undirected weighted $n$-vertex graph $G = (V,E,\omega)$, and a pair of parameters $\alpha \ge 1$ and $\beta = 1,2,\ldots$, a graph $G' = (V,H,\omega')$, $H \cap E = \emptyset$,  is called an
{\em $(\alpha,\beta)$-hopset} of $G$ if for every pair $u,v \in V$ of vertices we have
$$d_G(u,v) \le d_{G \cup G'}^{(\beta)}(u,v) \le \alpha \cdot d_G(u,v)~.$$
Here $\tG = G \cup G'$ is the union graph of $G$ and $G'$, i.e., $G \cup G' = (V,E \cup H,\tilde{\omega})$, where for every edge $e \in E$, $\tilde{\omega}(e) = \omega(e)$, and for every $e \in H$, $\tilde{\omega}(e) = \omega'(e)$.
Also,  $d_{\tilde{G}}^{(\beta)}$ stands for a {\em $\beta$-bounded} distance function in $\tilde{G}$, i.e., $d_{\tilde{G}}^{(\beta)}(u,v)$  is the length of the shortest $u-v$ path in $\tilde{G}$ that contains at most $\beta$ hops. The parameter $\beta$ is called the {\em hopbound} of the hopset $G'$, and $\alpha$ is called the {\em stretch} of the hopset.

Hopsets turn our to be extremely useful for exact and approximate distance-related computations in distributed, dynamic, parallel and streaming settings. They also constitute  fascinating combinatorial objects of independent interest.

Exact hopsets  (i.e., hopsets with $\alpha = 1$) were implicitly studied by Ullman and Yannakakis
\cite{UY91}, by Klein and Sairam \cite{KS93} and by Shi and Spencer \cite{SS99}.
{\em Near-exact} hopsets, i.e., hopsets with $\alpha = 1+\eps$, for an arbitrarily small $\eps > 0$, were introduced in a seminal paper by Cohen \cite{C94}. For an input undirected possibly weighted $n$-vertex graph, Cohen's algorithm constructs $(1+\eps,\beta)$-hopsets of size
$\tO(n^{1+1/\kappa})$, and with a polylogarithmic hopbound $\beta$.
Specifically, $$\beta_{Coh} = O\left({{\log n} \over \eps}\right)^{O(\log \kappa)}~.$$

Additional constructions of near-exact hopsets were given by Bernstein \cite{B09}, Henzinger et al. \cite{HKN14}, and by Miller et al. \cite{MPVX15}.  The hopset of \cite{B09} has hopbound
$O(\log n \cdot (1/\eps)^\kappa)$, and size $\tO(\kappa \cdot n^{1+1/\kappa} \cdot \log \Lambda)$, where $\Lambda$ is the aspect ratio of the graph.\footnote{The {\em aspect ratio} $\Lambda$ of a weighted graph $G = (V,E,\omega)$ is the ratio betwen $\max_{u,v} d_G(u,v)$ and
$\min_{u \neq v} d_G(u,v)$.}
The hopsets of Henzinger et al. \cite{HKN14,HKN16} have hopbound
$\mathit{exp}\{\tO_\eps(\sqrt{\log n \cdot \log\log n})\}$ and size
$n \cdot \mathit{exp}\{\tO_\eps(\sqrt{\log n \cdot \log\log n})\} \cdot \log^{O(1)} \Lambda$.
The hopsets of \cite{MPVX15} have hopbound $n^\gamma$, for an arbitrarily small constant $\gamma > 0$, and size $O(n)$.

The first construction of hopsets with {\em constant} hopbound (and non-trivial size guarantee) was given by the current authors in \cite{EN16}.  Specifically, we showed there that for any pair of parameters $\eps > 0$ and $\kappa = 1,2,\ldots$, there exists $\beta = \beta(\eps,\kappa) = \beta_{EP}$, such that for every undirected weighted $n$-vertex graph $G = (V,E,\omega)$, there exists a $(1+\eps,\beta)$-hopset with $O(n^{1+1/\kappa} \cdot \log n)$ edges. Note the striking similarity between this result and the result of Elkin and Peleg \cite{EP01} concerning near-additive spanners. Remarkably, not just the results are similar, but also their proofs are closely related. We will elaborate on this relationship below.

Interestingly, the same phenomenon occurs for the Thorup-Zwick's construction \cite{TZ06} of near-additive spanners and emulators as well. In \cite{EN19,HP17} it was shown that Thorup-Zwick's construction not only gives rise to universal near-additive emulators, but also to {\em universal} hopsets. Specifically, the algorithm of \cite{EN19,HP17,TZ06}, given an input $n$-vertex graph and a parameter $\kappa$, constructs a hopset of size $O(n^{1+1/\kappa})$, which serves as a $(1+\eps,\beta)$-hopset with $\beta = \beta_{EP}(\eps,\kappa) =
\left({{\log \kappa } \over \eps}\right)^{\log \kappa - 2}$, {\em simultaneously} for all $\eps > 0$.

%%%%%%%%%%%%%%%%%%%%%%%
\subsection{Discussion}
\label{sec:discussion}
%%%%%%%%%%%%%%%%%%%%%%%%%%%%%%

With these results in mind, it is natural to wonder why are near-additive spanners and near-exact hopsets that similar? After all, there are some apparent significant differences. First, near-additive spanners apply to unweighted graphs\footnote{Even though there are some results \cite{E01,EGN19} about weighted graphs as well.}, while near-exact hopsets apply to weighted graphs. Second, the meaning of the parameter $\beta$ is very different. For spanners it is the additive error term, while for hopsets it is the hopbound. Third, spanners are subgraphs of the input graph, while hopsets are sets of edges that are  not present in the original graph.
(This distinction becomes blurred if one considers emulators instead of spanners. Nevertheless, an emulator, like a spanner, is used on its own, while hopset is used together with the edges of the original graph.)

We will next shortly discuss these techniques for spanners and hopsets. In the technical part of this survey we will sketch proofs of these results, and highlight the similarities and differences  between them.

As was discussed above, there are three main approaches to building near-additive spanners and near-exact hopsets. The first one is the superclustering and interconnection approach, which was introduced by  \cite{EP01} in the context of near-additive spanners, and used  in \cite{EN16} for constructing hopsets. The second one, closely related to the first one, is the universal extension of the superclustering and interconnection approach. It was introduced by \cite{TZ06} in the context of  spanners and emulators, and used in \cite{EN19,HP17} in the context of hopsets.
The third one, based on neighborhood  covers, was originated in Cohen's construction of hopsets \cite{C94}. It was later used in \cite{E01,EZ04} for building near-additive spanners.
Because of space considerations, we will  focus on the first two approaches in this survey.

The superclustering and interconnection approach, which we describe in detail in Section \ref{sec:supinterconn}, works roughly as follows. It proceeds for $\ell = \log \kappa$ phases, indexed $0,1,\ldots,\ell-1$. (Throughout the survey, we assume, for simplicity, that $\log \kappa = \log_2 \kappa$ is an integer. This has only a very minor impact on the cited bounds.) On phase 0, its input partition $\cP_0 = \{\{v\} \mid v \in V\}$ is the collection of singletons. One uses a sequence of degree thresholds, $\deg_0,\deg_1,\ldots,\deg_{\ell-1}$, the simplest of which is $\deg_i = n^{2^i \over \kappa}$ \cite{EP01}, and a sequence of distance thresholds $\delta_i = (1/\eps)^i$. Each cluster $C$ of the input partition $\cP_i$ that has at least $\deg_i$ other clusters of $\cP_i$ at distance at most $\delta_i$ from it, becomes {\em superclustered}, i.e., merged into a next-level cluster, a cluster of $\cP_{i+1}$.  Spanning trees of superclusters are added into the spanner/hopset. (On phase $\ell-1$, the superclustering step is skipped, and the algorithm proceeds directly to the interconnection step.)

In \cite{EP01}, this is done directly. Specifically, one initializes the set $\cU_i$ of uncovered clusters as $\cP_i$. Then one iteratively finds such clusters $C \in \cP_i$ with many uncovered nearby clusters, creating superclusters around them, and marking them as covered.
 For reasons of efficiency and parallelism, in \cite{EN16,EN17}, this is done by sampling clusters of $\cP_i$ with probability ${1 \over {\deg_i}}$, and creating superclusters around the sampled clusters.
In \cite{EM19} the same step is performed by computing ruling sets.

At any rate, once we are done with superclustering, we move to the interconnection step. On this step pairs of clusters are interconnected, i.e., shortest paths between them  (or direct edges, in the case of hopsets/emulators) are inserted into the spanner (respectively, hopset/emulator).

The stretch analysis of this construction considers a pair $u,v \in V$ of vertices, and a shortest
path $\pi(u,v)$ between them.  The path is partitioned into segments of length
$\delta_{\ell-1} = (1/\eps)^{\ell-1}$. On each such a segment $x-y$ one identifies the leftmost and the rightmost $\cP_{\ell-1}$ clusters $C_L$ and $C_R$. The substitute spanner's path $\pi'(x,y)$ that the stretch analysis finds uses a direct $C_L-C_R$ shortest path. The latter was inserted into the spanner on the $(\ell-1)$st phase, because $d_G(C_L,C_R) \le \delta_{\ell-1}$.
Then the stretch analysis zooms in into the $x-C_L$ subpath, and into the $C_R-y$ subpath. Both these subpaths are free of $\cP_{\ell-1}$ clusters, and the stretch analysis exploits this to provide small-stretch substitute spanner's paths for them.

We note that the radii of $C_L$ and $C_R$ are both $O((1/\eps)^{\ell-2})$, i.e., by one order of magnitude (that is, by a factor of $1/\eps$) smaller than the length of the segment $x-y$.
The same phenomenon occurs on lower levels of stretch analysis as well, i.e., in segments of the subpaths $x-C_L$ and $C_R-y$. These segments are of length $(1/\eps)^{\ell-2}$, while the maximum radius of a cluster that appears on these segments is $O((1/\eps)^{\ell-3})$, etc.
Hence, roughly speaking, we have  multiplicative stretch of $1+\eps$ on every level of stretch analysis, and thus the overall stretch is $1 + O(\eps \cdot \ell)$. The additive error term stems from the fact
that $d_G(u,v)$ might be shorter than $\delta_{\ell-1} = (1/\eps)^{\ell-1}$. In this case one would not be able to charge the radii of $C_L$ and $C_R$ to the length of the segment $x-y$,
and the additive term accounts for that.

The construction of hopsets that employs the superclustering and interconnection method \cite{EN16} proceeds along very similar lines. In its simplest form it builds a separate hopset $H_j$  for each distance scale $[2^j,2^{j+1})$, for $j = 0,1,\ldots,\lceil \log \Lambda \rceil$.
For each fixed $j$, the hopset $H_j$ takes care of pairs $u,v \in V$ of vertices with $d_G(u,v) \in [2^j,2^{j+1})$. The ultimate hopset is $H = \bigcup_{j=0}^{\lceil \log \Lambda \rceil} H_j$.

We then define a distance unit $\gamma = {{2^j} \over {(1/\eps)^{\ell-1}}} = \eps^{\ell-1} 2^j$. (As we aim at hopbound of $(1/\eps)^{\ell-1}$, one can assume that $2^j \ge (1/\eps)^{\ell-1}$.) Then the distance thresholds $\delta_i$ are defined as $\gamma \cdot (1/\eps)^i$, i.e., in the same way as for near-additive spanners, except for scaling by a factor of $\gamma$.
We then conduct the same superclustering and interconnection algorithm as for the spanner's construction, with the same degree thresholds, and distance thresholds $\delta_i$ as above.
The only difference is that instead of inserting shortest paths between pairs of vertices $z,z'$ into the spanner, here we insert direct hopset edges $(z,z')$ of weight $d_G(z,z')$. (This also happens in the construction of emulators.)

 The stretch analysis of the resulting hopset is also conducted very similarly to the case of spanners. There are some technicalities that have to do with the fact that we deal with weighted graphs in the case of hopsets, and thus we may not be able to partition the shortest path into segments of length precisely $\delta_{\ell-1} = (1/\eps)^{\ell-1} \gamma$. However, one can easily overcome these difficulties. (See Section \ref{sec:sp_anal}.)
Another difference is that one can use edges of the original graph in the substitute path in the case of hopsets, while this is not the case for spanners. This actually makes the stretch analysis easier in the former case. Finally, in the case of hopsets
one also needs to carefully analyze the number of hops used in
the substitute path. Intuitively, the shortest path $\pi(u,v)$ is partitioned to $\approx (1/\eps)^{\ell-1}$ subsegments of length $\gamma$, and for each of them $O(1)$ hops suffice.
Thus, the hopbound is, up to rescaling of $\eps$, equal to  $O((1/\eps)^{\ell-1})$.

Next we overview the construction of Thorup-Zwick's emulators \cite{TZ06} and hopsets of \cite{EN19,HP17}, while focusing on their relationship to the superclustering and interconnection method. As was already mentioned, these emulators and hopsets can be viewed as a scale-free version of
spanners and hopsets from \cite{EP01,EN16}.

the algorithm of \cite{TZ06} works as follows. It defines $A_0 = V$, and for $i = 0,1,\ldots,\ell-1$, vertices of $A_{i+1}$ are obtained from those of $A_i$ by sampling each $v \in A_i$
independently
with
probability ${1 \over {\deg_i}}$. The sequence $\deg_0,\deg_1,\ldots,\deg_{\ell-1}$ of degree thresholds is defined exactly as in \cite{EP01}.

Given this hierarchy of subsets $V = A_0 \supseteq A_1 \supseteq \ldots \supseteq A_{\ell-1}$, the algorithm defines for every vertex $v \in A_i$, $i < \ell-1$, its {\em pivot} $p(v)$ to be
the closest $A_{i+1}$-vertex to $v$. (Ties are broken arbitrarily but consistently.)
We also define the {\em bunch} of $v \in A_i$ by
$$\Bunch(v) =  \{u \in A_i \mid d_G(v,u) < d_G(v,A_{i+1})\}~.$$
It is the set of all vertices of $A_i$ that are closer to $v$ than the pivot of $v$.
For any $v \in A_{\ell-1}$, its bunch is defined as the entire $A_{\ell-1}$.

The algorithm then inserts into the emulator $H$ (and into the hopset)
the edges
$\bigcup_{i=0}^{\ell-1} \{(v,u) \mid v \in A_i, u \in \Bunch(v)\}$, and also the edges
$\bigcup_{i=0}^{\ell-2} \{(v,p(v))\}$.
This completes the description of the construction. Intuitively, the edges
$\bigcup_{i=0}^{\ell-2} \{(v,p(v))\}$ are {\em superclustering} edges, i.e., edges that connect an $i$-level cluster center to its $(i+1)$st level parent. The edges of
$\bigcup_{i=0}^{\ell-1} \{(v,u) \mid v \in A_i, u \in \Bunch(v)\}$ are {\em interconnection} edges, i.e., edges that connect pairs of cluster centers of the same level.

For the stretch analysis, we consider a pair $u,v \in V$ of vertices, at distance $d = d_G(u,v)$ from one another. If we analyze $H$ as an emulator,  then we partition a shortest path $\pi(u,v)$ between them into segments
of length $\delta_{\ell-1} = (1/\eps)^{\ell-1}$. (Recall than $\eps > 0$ is not a parameter of the algorithm in these scale-free constructions. Rather, it is a parameter of the {\em analysis}, which applies for any $\eps > 0$.) These segments are then partitioned into $1/\eps$ subsegments
of length $\delta_{\ell-2}$, and those are again partitioned to $1/\eps$ subsegments, up until we reach single edges. (To analyze $H$ as a hopset, we partition $\pi(u,v)$ of length $d$ into
$\approx 1/\eps$ segments of length $\approx d \cdot \eps$, and each of them into $\approx 1/\eps$ segments of length $\approx d \cdot \eps^2$, up until the bottom level, where each subsegment has length $\approx d \cdot \eps^{\ell-1}$.)

Now a segment $x-y$ of level $i$ (i.e., of length $\delta_i$) is called {\em successful}, if it admits a substitute path of length $1 + O(i \cdot \eps)$ between its endpoints in the emulator. For an unsuccessful segment, it can be argued that its  endpoints $x,y$ admit  nearby $(i+1)$st pivots $x',y'$, respectively. This is argued by an induction on $i$. The base case follows from definitions of pivots and bunches (with stretch 1). For the induction step, the analysis considers $i$-level subsegments
of an $(i+1)$st level segment. If they are all successful, then we get a stretch of $1 +O(i \cdot \eps)$ for the entire segment, and we are done. Otherwise, we consider the leftmost and the rightmost unsuccessful subsegments $x_L-y_L$ and $x_R-y_R$. (The case that there is just one unsuccessful subsegment is even simpler. See Section \ref{sec:scalefree}.) By the induction hypothesis, there are
$(i+1)$st level pivots $z_L$, $z_R$, with $z_L$ being close to $x_L$ and $z_R$ to $y_R$.
Now either $z_R \in \Bunch(z_L)$, and so the edge $(z_L,z_R)$ is in the emulator. We then obtain a short substitute emulator's $x-y$ path, that consists of the subpaths $x-x_L$, $x_L-z_L$,
$z_L-z_R$, $z_R-y_R$, and finally, $y_R-y$.
Otherwise, there is a nearby $(i+1)$nd pivot $z$ to $z_L$, which, by triangle's inequality, is also close to $x$. This completes the inductive proof.

This inductive statement is used with $i=\ell-1$. At this level all segments are successful, just because $A_\ell = \emptyset$. Hence the emulator provides stretch $1+O(\eps \cdot \ell)$. In
the case of hopsets one needs also to carefully count the number of hops, but other than that the stretch analysis proceeds along the same lines.

%%%%%%%%%%%%%%%%%%%%%
\subsection{Organization}
%%%%%%%%%%%%%%%%%%

In Section \ref{sec:supinterconn} we describe the superclustering and interconnection approach in more detail.
In Section \ref{sec:scalefree} we do so for its scale-free extension.

%%%%%%%%%%%%%%%%%%%%%%%%%%
\section{Superclustering and Interconnection}
\label{sec:supinterconn}
%%%%%%%%%%%%%%%%%%%%%%%%%%%%%

This section is devoted to the superclusterig and interconnection method of constructing near-additive spanners and hopsets \cite{EP01,EN16,EM19}.  We start (Section \ref{sec:sp}) with describing the construction of near-additive spanners, and then proceed (Section \ref{sec:hop}) to hopsets.

%%%%%%%%%%%%%%%%%%%%%%%%%%%%
\subsection{Spanners}
\label{sec:sp}
%%%%%%%%%%%%%%%%%%%%%%%%%%%%%

%%%%%%%%%%%%%%%%%%%%%%%%%%%%%%%
\subsubsection{Algorithm}
\label{sec:sp_alg}
%%%%%%%%%%%%%%%%%%%%%%%%%%%%%%%%%%%

Let $Q$ be the {\em ground} partition of the graph $G = (V,E)$, i.e., $Q = \{C_1,C_2,\ldots,C_q\}$, for some integer $q \ge 1$, is a collection of pairwise disjoint clusters such that $V = \bigcup_{C \in Q} C$. Moreover, each cluster $C \in Q$ has a designated center $r_C$, and the {\em radius}
of the partition, i.e., the maximum radius of one of its clusters (with respect to its designated center) is
$$\mathit{Rad}(C) = \max_{u \in C} \{d_{G(C)}(r_C,u)\} \le \kappa-1~,$$
for a parameter $\kappa$ that controls the stretch-size tradeoff of the ultimate spanner.

The {\em supergraph} $\cG = (Q,\cE)$ induced by the ground partition is defined by
$$\cE = \{(C,C') \mid C \neq C', C,C' \in Q, \exists (v,v') \in E, v \in C,v' \in C'\}~.$$
The ground partition $Q$ has the property that the supergraph $\cG$ is sparse, i.e., $|\cE| = O(n^{1+1/\kappa})$.

Moreover, the ground partition $Q$ has the  property that $Q = \bigcup_{i=0}^{\kappa-1} Q_i$, where all clusters in $Q_i$ have radius $i$, contain at least $n^{i/\kappa}$ vertices each, and have at most $n^{(i+1)/\kappa}$ ``outgoing'' neighbors (so that the total number of neighboring clusters is $O(n^{1+1/\kappa})$).

For convenience, we will assume that
$\kappa$ is of the form $\kappa = 2^\ell - 1$, for an integer $\ell$.
It is easy to adapt the constructions to the case of a general integer parameter $\kappa$.
We partition the set of indices $\{0,1,\ldots,\kappa-1\}$ into subsets
$\{0\},\{1,2\},\{3,4,5,6\},\ldots,\{2^{\ell-1} - 1,2^{\ell-1},\ldots,2^{\ell}-2\}$.
Let $\hQ_0 = Q_0$, $\hQ_1 = Q_1 \cup Q_2, \ldots$, and
 $\hQ_{\ell-1}=Q_{2^{\ell-1}-1} \cup \ldots \cup Q_{2^{\ell}-2}$.

Constructions of such ground partitions are well-known, and can be found, e.g., in \cite{PS89,AP92,HZ96,EP01}.
We note that modern constructions of near-additive spanners \cite{EN17,EM19} that follow the superclustering and interconnection paradigm manage to bypass ground partitions altogether. However, the original construction of \cite{EP01} that does use them is somewhat simpler.

The spanner $H$ is initialized to contain the union of BFS spanning trees of all clusters $C$ of the ground partition $Q$.
For each cluster $C \in Q$, the BFS tree is rooted in its designated center $r_C$.  We also insert into the spanner one edge $e =(u,v)$ for each pair of neighboring clusters $C,C' \in Q$ (i.e., e.g., $u\in C$, $v \in C'$). The overall number of edges inserted to the spanner so far is $O(n^{1+1/\kappa})$. (See \cite{PS89,AP92,HZ96,EP01}.)

The algorithm itself proceeds for $\ell$ phases.
%(Recall that $\kappa = 2^\ell-1$.)
At the beginning of each phase $i$, $0 \le i \le \ell -1$, we have the input partition $\cP_i$ of clusters.  For $ i \le \ell-2$,
the phase $i$ {\em superclusters} some of these clusters into larger clusters (aka {\em superclusters}).
The resulting partition $\chP_i$, union with the collection $\hQ_{i+1}$ of clusters from the ground partition, forms the input for the next phase $i+1$. (On phase 0, the input is $\cP_0 = \hQ_0 = Q_0$.)
Some other clusters of $\cP_i$ are not involved in superclustering. The set of these clusters is called $\cU_i$, the set of unsuperclustered clusters of phase $i$. On the last phase $i = \ell-1$, the superclustering step is omitted, and we define $\cU_{\ell-1} = \cP_{\ell-1}$.

On all phases $i=0,1,\ldots,\ell-1$, the unsuperclustered clusters (the set $\cU_i$) of this phase proceed to the {\em interconnection}  step. On the interconnection step shortest paths between nearby clusters of $\cU_i$ are added into the spanner $H$.  An invariant of the algorithm is that each of the superclusters of $\cP_i$ has size at least $n^{{2^i - 1} \over \kappa}$, for all $i = 0,1,\ldots,\ell-1$. At the beginning of phase $i$, the partition $\cP_i$ is created as a union of $\chP_{i-1}$ (the output of phase $i-1$)  with $\hQ_i$. Recall that each of the clusters of $\hQ_i$ has size at least $n^{{2^i - 1} \over \kappa}$ as well.

The algorithm also employs sequences of degree and distance thresholds. On each phase $i$, it uses the parameters
$\deg_i = n^{{2^i} \over \kappa}$ as {\em degree threshold} and $\delta_i = (1/\eps)^i$ as {\em distance threshold}.

Next, we take a closer look on each particular phase $i = 0,1,\ldots,\ell-2$. (The last phase $i = \ell-1$ is slightly different, as it has no superclustering step.)
The algorithm checks if there exists a cluster $C \in \cP_i$, such that at distance at most $\delta_i$ (in $G$)  from $C$, there are at least $\deg_i$ uncovered clusters $C' \in \cP_i$. (Initially all clusters of $\cP_i$ are uncovered, i.e., $\cU_i$ is initialized as $\cP_i$.) If there is such a cluster $C$, then the algorithm creates a supercluster $\hC$ around it, centered at $C$. This supercluster includes all the other uncovered clusters $C' \in \cP_i$ at distance at most $\delta_i$
(in $G$) from $C$.  Shortest paths
between $C$ and each of these clusters $C'$
 are added to the spanner $H$. Finally, $C$ and all these clusters $C'$ that are merged into $\hC$ are removed from $\cU_i$, i.e., they are marked as covered. Then the algorithm iterates. The resulting collection of superclusters $\hC$
created in this way is the aforementioned set $\chP_i$. Together with $\hQ_{i+1}$ it constitutes the collection $\cP_{i+1}$, which serves as input to phase $i+1$. This concludes the superclustering step of phase $i$.

On the interconnection step of phase $i$, each pair of clusters of $\cU_i$ that are at distance at most $\delta_i$ from one another in $G$, are interconnected with one another via shortest paths. These shortest paths are added into the spanner $H$.

The last phase $i = \ell-1$ is special, because the overall number of clusters in the input collection $\cP_{\ell-1}$ is at most $O(n^{{\kappa+1} \over {2\kappa}})$. Consequently, one interconnects all pairs of nearby clusters of $\cP_{\ell-1}$ (i.e., clusters at distance at most $\delta_{\ell-1}$ from one another in $G$).
%As was mentioned above, the superclustering step is skipped on phase $\ell-1$.

This concludes the description of the algorithm.

%%%%%%%%%%%%%%%%%%%%%%%%%%%%%
\subsubsection{Analysis}
\label{sec:sp_anal}
%%%%%%%%%%%%%%%%%%%%%%%%%%%%%%%%%

We next sketch its analysis.

{\bf Size:} We start with the size analysis. The number of edges inserted into $H$ during the initialization step is, as was mentioned above, $O(n^{1+1/\kappa})$. This follows from properties of the ground partition \cite{PS89}.

Consider some fixed phase $i = 0,1,\ldots,\ell-1$.  It can be easily seen inductively that each cluster $C \in \cP_i$ has size at least $n^{{2^i - 1} \over \kappa}$. Since they are disjoint, we conclude that $|\cP_i| \le n^{1 - {{2^i -1} \over \kappa}}$. Recall that $\deg_i = n^{{2^i} \over \kappa}$. Hence the number of paths inserted into the spanner by the interconnection step of phase $i$ is at most $|\cP_i| \cdot \deg_i \le n^{1+1/\kappa}$.
Each path contains at most $\delta_i$ edges (because we connect nearby clusters). Thus, the number of edges inserted on this step is $O(\delta_i \cdot n^{1+1/\kappa})$.

Consider the superclustering step (for $i < \ell -1$). Let $\hC_1,\hC_2,\ldots,\hC_p$ be the set of created superclusters. For each index $j \in [p]$, let $C_j$ be the center cluster of $\hC_j$, i.e., the cluster around which the supercluster $\hC_j$ was created.  Also, let $C_{j1},C_{j2},\ldots,C_{j(q_j)}$ denote the other clusters superclustered into the supercluster $\hC_j$. Then the collection of edges
$$\{(C_j,C_{j1}),(C_j,C_{j2}),\ldots,(C_j,C_{j(q_j)})\} \mid 1 \le j \le p\}$$ forms a forest (a collection of disjoint stars), and thus  contains less than $n$ edges. For each of these edges, at most $\delta_i$ edges of the respective shortest path between $C_j$, for some $1 \le j \le p$, and some $C_{jh}$, for some $1 \le h \le q_j$, are inserted into the spanner $H$.
Thus,  the superclustering step of phase $i$ inserts into the spanner $O(\delta_i \cdot n)$ edges. Thus, altogether phase $i$ inserts into the spanner $O(\delta_i \cdot n^{1+1/\kappa})$ edges.
Hence overall
$$|H| = O(n^{1+1/\kappa}) \cdot \sum_{i=0}^{\ell-1} \delta_i = O_{\eps,\kappa}(n^{1+1/\kappa})~.$$
(Recall that $\delta_i = (1/\eps)^i$.)

{\bf Stretch:}
Next we outline the stretch analysis of this construction.

Note that $\cU_0 \cup \cU_1 \cup \ldots \cU_{\ell-1}$ is a partition of $V$.
Let $\cU^{(i)} = \bigcup_{j=0}^i \cU_j$, for all $i \in [0,\ell-1]$.
Observe also that all clusters of $\cU_0$ are singletons, i.e., their radius (denoted $R_0$) is 0. Each of the clusters $C$ in $\cU^{(\ell-1)}$ has a designated center $r$, and the {\em radius} of $C$, denoted $\Rad(C)$, is defined as
$\max_{v \in C} d_H(r,v)$.

\begin{figure}
	\begin{center}
		\includegraphics[width=0.7\textwidth]{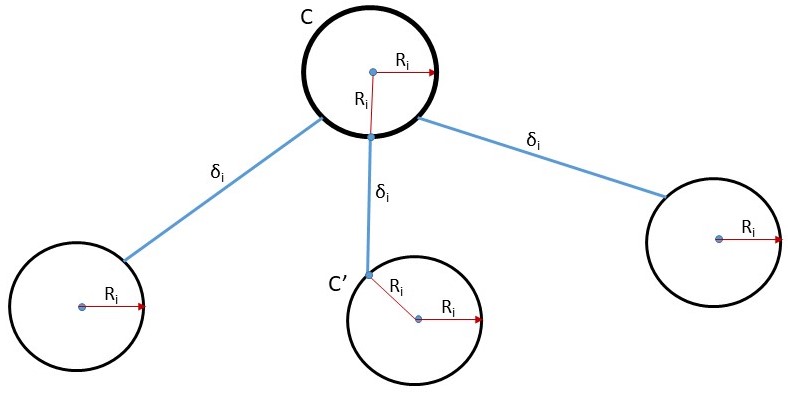}
		\caption{\small A drawing justifying the inequality $R_{i+1}\le 3R_i + \delta_i$, where $R_i$ is the radius of a level $i$ cluster, and $\delta_i$ is the bound on the search distance at level $i$. }
		\label{fig:radius}
	\end{center}
\end{figure}

Generally, observe that the radius $R_{i+1}$ of a cluster $\hC \in U_{i+1}$, for some $0 \le i \le \ell-2$, is at most
$3R_i + \delta_i = 3R_i + (1/\eps)^i$. (Here $R_i$ is the maximum radius of a cluster in $\cU_i$.)
See Figure \ref{fig:radius} for an illustration.

Thus, $R_1 = 1$, and generally, for $0 \le i \le \ell-2$, we have
$$R_{i+1} = \sum_{j=0}^i 3^{i-j}(1/\eps)^j < 2 \left({1 \over \eps}\right)^i~,$$
assuming $\eps < 1/6$.

Let $u,v \in V$ be a pair of vertices, and let $\pi(u,v)$ be a fixed shortest $u-v$ path in $G$. We partition it into segments of length $(1/\eps)^{\ell-1}$, except the last segment that may be shorter.
Consider a particular fixed segment $x-y$ of this path, of length at most $(1/\eps)^{\ell-1} = \delta_{\ell-1}$.

It is convenient to imagine the path $\pi(u,v)$ and the subpath $\pi(x,y)$  as going from left to right, with $u$ and $x$ being the left endpoints and $v$ and $y$ being the right endpoints of their respective paths.
Let $z$ and $w$ be the leftmost and the rightmost $\cU_{\ell-1}$-clustered vertices on $\pi(x,y)$.
(We assume that they exist. It is possible that $z = w$. If no such a vertex exists, the analysis is actually simpler, as will be further indicated below.)
Let $C_z,C_w \in \cU_{\ell-1}$ be the clusters of $z$ and $w$, respectively, i.e., $z \in C_z$, $w \in C_w$.

Then $d_G(C_z,C_w) \le d_G(z,w) \le (1/\eps)^{\ell-1}$, and thus, a shortest path $\tilde{\pi}$ between $C_z$ and $C_w$ was inserted into the spanner $H$.  It follows that there exist vertices $\tilde{z} \in C_z$, $\tilde{w} \in C_w$, such that $\tilde{\pi}$ is the shortest $\tilde{z}-\tilde{w}$ path. Since spanning trees of radius at most $R_{\ell-1}$ for each cluster $C \in \cU_{\ell-1}$ are contained in the spanner $H$, we conclude that
$$d_H(z,w) \le d_G(C_z,C_w) + 4 \cdot R_{\ell-1} \le (1/\eps)^{\ell-1} + 8 \cdot (1/\eps)^{\ell-2}~.$$
Let $z'$ (respectively, $w'$) be the left (resp., right) neighbor of $z$ (resp., $w$) on $\pi(x,y)$, if exists.
Since $z'$ and $z$ belong to neighboring clusters of the ground partition $Q$, there is a path in $H$ of length at most $4(\kappa-1)+1$ between them.
Thus, $$d_H(z',w') \le (1/\eps)^{\ell-1} + 8 \cdot ((1/\eps)^{\ell-2} + \kappa)~.$$
For simplicity, we suppress the term $\kappa$ in this expression, as it is dominated by
$(1/\eps)^{\ell-2} = (1/\eps)^{\log (\kappa+1) - 2}$.

So the overall overhead so far that the spanner's path incurs in comparison to the original shortest path is
$O((1/\eps)^{\ell-2})$, for each segment of length $(1/\eps)^{\ell-1}$. This amounts to the multiplicative stretch of $1+ O(\eps)$.  The last segment of the path $\pi(u,v)$, which may be of length smaller than $(1/\eps)^{\ell-1}$, is responsible for the additive stretch of $O((1/\eps)^{\ell-2})$.

But we are not yet done. The spanner's path still needs to reach from $x$ to $z'$ and from $w'$ to $y$.  Observe that both these subsegments contain only vertices clustered at $\cU^{(\ell-2)}$ (i.e., they are not clustered in $\cU_{\ell-1}$).\footnote{The case that the entire $x-y$ path $\pi(x,y)$ has no $\cU_{\ell-1}$-clustered vertices is actually a special case of the case considered here.}
We partition each of them into subsegments of length $\delta_{\ell-2}= (1/\eps)^{\ell-2}$ each, except the last subsegment that may be shorter.

On each such a subsegment $x'-y'$, we find the leftmost and the rightmost $\cU_{\ell-2}$-clustered vertices
$z_{\ell-2}$ and $w_{\ell-2}$. Let $z'_{\ell-2}$ be the left neighbor of $z_{\ell-2}$, and $w'_{\ell-2}$ be the right neighbor of $w_{\ell-2}$ along the path. The respective clusters
$C(z_{\ell-2}),C(w_{\ell-2}) \in \cU_{\ell-2}$ containing $z_{\ell-2}$ and $w_{\ell-2}$, respectively, have radius at most $R_{\ell-2} \le 2(1/\eps)^{\ell-3}$. Hence an analogous computation to the one we did above for the $z-w$ path shows that the spanner $H$ contains a $z'_{\ell-2}-w'_{\ell-2}$ path of length at most $d_G(z'_{\ell-2},w'_{\ell-2})  +
8 \cdot((1/\eps)^{\ell-3} + \kappa/2)$.
(The second term is $\kappa/2$ and not $\kappa$, because clusters of the ground partition that may end up in a $\cU_{\ell-2}$
cluster
belong to $\hQ_{\ell-2}$, and thus their radii are at most $\kappa/2$.)
In other words, on each subsegment of length $(1/\eps)^{\ell-2}$, the spanner's path pays an overhead of
$8 \cdot((1/\eps)^{\ell-3} + \kappa/2)$, i.e., another multiplicative factor of $1 + O(\eps)$.

We then proceed by zooming in into subsegments between $x'$ and $z'_{\ell-2}$, and between $w'_{\ell-2}$ and $y'$,
They are $\cU^{(\ell-3)}$-clustered, and thus analogous considerations can be applied to them.
Ultimately, this stretch analysis accumulates an overhead of $O(\eps)$-fraction of the length of the original path for $\ell$ times, leading to an overall multiplicative stretch of $1 +O(\eps \cdot \ell)$.
The additive error of the spanner manifests itself on the last segment $x-y$ of the partition of $\pi(u,v)$ into segments of length $\delta_{\ell-1} = (1/\eps)^{\ell-1}$ is of length much smaller than $(1/\eps)^{\ell-1}$.
Then the spanner's  path pays an overhead of $O(R_{\ell-1}) = O((1/\eps)^{\ell-2})$, and this overhead cannot be charged to edges of the segment $x-y$, because the latter segment is too short.

To summarize, the spanner provides a stretch of $(1 + O(\eps \cdot \ell),(1/\eps)^{\ell-2})$. By rescaling, i.e.,
setting $\eps' = O(\eps \cdot \ell)$, one obtains a $(1 +\eps',O\left({\ell \over {\eps'}}\right)^{\ell-2})$-spanner.
Hence we have additive term  $$\beta = O\left({{\log \kappa} \over {\eps'}}\right)^{\log (\kappa+1)-2}~.$$

We conclude this section with the following theorem:

\begin{theorem} \cite{EP01}
\label{thm:sp_supinterconn}
For every pair of parameters $\eps > 0$ and $\kappa = 1,2,\ldots$, there exists $\beta = \beta(\eps,\kappa)
= O\left({{\log \kappa} \over \eps}\right)^{\log (\kappa+1) - 2}$, such that for every unweighted undirected $n$-vertex
 graph $G = (V,E)$ there exists a $(1+\eps,\beta)$-spanner with $O_{\eps,\kappa}(n^{1+1/\kappa})$ edges.
\end{theorem}

We note that if one is interested in an {\em emulator} as opposed to spanner, one can use the very same construction,
but every time it inserted a shortest path between a pair of clusters $C,C'$ into the spanner, the emulator will include
a weighted edge between their respective centers $r_C$ and $r_{C'}$ of weight equal to the distance $d_G(r_C,r_{C'})$ between these centers. It is easy to verify that the resulting emulator will have size $O_\kappa(n^{1+1/\kappa})$ (as opposed to $O_{\eps,\kappa}(n^{1+1/\kappa})$), i.e., its size will no longer depend on $\eps$. One can also obtain this property for spanners constructed via the superclustering and interconnection approach, but via  a slightly more involved construction that involves distance preservers \cite{CE05}, and with a slightly inferior additive error $\beta$ \cite{EN17}.

%%%%%%%%%%%%%%%%%%%%%%%%
\subsection{Hopsets}
\label{sec:hop}
%%%%%%%%%%%%%%%%%%%%%%%%%%

In this section we argue that the same approach of superclustering and interconnection can be used to produce hopsets, with parameters similar to those of the corresponding near-additive spanners.

%%%%%%%%%%%%%%%%%%
%\subsubsection{The Algorithm}
%%%%%%%%%%%%%%%%%%

The algorithm produces a separate hopset for each distance scale. Assume that the smallest  edge weight is  1,
and let the aspect ratio $\Lambda$ denote the maximum distance between a pair of vertices $u,v$ in the input weighted undirected graph $G = (V,E,\omega)$.
Our ultimate hopset $H$ will be the union of single-scale hopsets $H_i$, where for each $i = 0,1,\ldots,\lambda = \lceil \log_2 \Lambda \rceil$, $H_i$ is the hopset that takes care of pairs of vertices $u,v$ with $d_G(u,v) \in [2^i,2^{i+1})$. (The last scale will always contain pairs with distance exactly $\Lambda$ as well.)
In \cite{EN16} we showed that one can get rid of the dependence on the aspect ratio in the size of the ultimate hopset
$H$. Here, however, we will describe a simpler construction in which $|H_i| = O_\kappa(n^{1+1/\kappa})$,\footnote{Specifically, $|H_i| = O(\log \kappa \cdot n^{1+1/\kappa})$. One can also eliminate the leading factor of $\log \kappa$ \cite{EN16}.} for every scale $i \in [0,\lambda]$, and thus,
$|H| = O_\kappa(\log \Lambda \cdot n^{1+1/\kappa})$.

We fix a scale $i$, denote $R = 2^i$, and construct a hopset $H_i$ that takes care of pairs of vertices $u,v$ with
$d_G(u,v) \in [R,2R)$. From now on it will be referred to as $H' = H_i$.

%Similarly to the construction os near-additive spanners from Section \ref{sec:sp_alg},
We initialize $\cP_0 = \{\{v\} \mid
 v \in V\}$ as the partition of $V$ into singletons. (In \cite{EN17} it was shown one can start with a partition into singletons when building near-additive spanners as well.)
We initialize the set of uncovered clusters as $\cU_0 \leftarrow \cP_0$.
%We then pick an uncovered cluster
%$C = \{v\} \in \cU_0$, such that at
%distance at most $\delta_0 = R \cdot \eps^{\ell-1}$ from it there are at most $\deg_0 = n^{{2^0} \over \kappa} = %n^{1/\kappa}$ uncovered clusters.
Let $\delta_0 = R \cdot \eps^{\ell-1}$.
 Generally, we define $\deg_i = n^{{2^i} \over \kappa}$, for all $i \in [0,\ell-1]$, exactly as in the construction of near-additive spanners in Section \ref{sec:sp_alg}. Also, we set
$\delta_i = \delta_0/\eps^i$, for all $i \in [0,\ell-1]$. In particular, $\delta_{\ell-1} = R$.
The way to think of it is that $\delta_0 = R \cdot \eps^{\ell-1}$ is the ``distance unit" of the construction. Scaling down by the distance unit, one obtains the same sequence of distance thresholds as in Section \ref{sec:sp_alg}.

Returning to the superclustering step of phase 0, if the algorithm finds an uncovered cluster $C \in \cU_0$ with at least $\deg_0 = n^{1/\kappa}$ other uncovered clusters $C' \in \cU_0$ at distance at most $\delta_0$ from it in $G$, then it creates a supercluster $\hC$ out of them centered at the center $r_C$ of $C$. (For a singleton cluster $C = \{v\}$, the center $r_C$  is set as $v$.) The supercluster is created by adding into it $C$, and the nearby clusters $C' \in \cU_0$ (at distance at most $\delta_0$ from $C$ in $G$). One also adds to the hopset the edges
$\{ (r_C,r_{C'}) \mid C' \in \hC\}$, with weights $\omega((r_C,r_{C'})) = d_G(r_C,r_{C'})$.
The cluster $C$ and the clusters $C'$ as above are then removed from $\cU_0$, i.e., they are marked as covered.
We then proceed to constructing the next supercluster in the same manner. The superclustering step (of phase 0) proceeds iteratively up until no additional supercluster can be created.

 The set $\cU_0$ of remaining {\em unclustered} clusters is then the input to the interconnection
step of phase 0. On this step each pair of nearby clusters $C,C' \in \cU_0$ (i.e., $d_G(C,C') \le \delta_0$) is interconnected by a direct hopset edge $(r_C,r_{C'})$ between their respective centers. Its weight is also set as $d_G(r_C,r_{C'})$, This concludes the description of phase 0.

The resulting collection $\cP_1$ of superclusters created on phase 0 is the input for phase 1. Phase 1 proceeds in the same way (as phase 0), except that its degree and threshold parameters are $\deg_1 = n^{{2^1} \over \kappa}$ and
$\delta_1 = \delta_0/\eps^1$. This is also the case for phases $i \ge 1$, that have $\deg_i = n^{{2^i} \over \kappa}$ and $\delta_i = \delta_0/\eps^i$. When the algorithm reaches phase $\ell-1$, all clusters of $\cP_{\ell-1}$ are already of the size at least $n^{{2^{\ell-1} - 1} \over \kappa} = n^{{\kappa-1} \over {2\kappa}}$. (By the same argument as in Section \ref{sec:sp_anal}.)  Hence $|\cP_{\ell-1}| \le n^{{\kappa+1} \over {2\kappa}}$ (because the clusters are disjoint). So the superclustering step of phase $\ell-1$ is skipped. Instead the algorithm proceeds directly to interconnecting all pairs of clusters of $\cP_{\ell-1}$. (We also set $\cU_{\ell-1} = \cP_{\ell-1}$, to reflect the intuition that all clusters of phase $\ell-1$ are uncovered.) By ``interconnecting" a pair $(C,C')$ of clusters, we again mean inserting into the hopset the edge $(r_C,r_{C'})$ between their respective centers, with weight $\omega((r_C,r_{C'})) = d_G(r_C,r_{C'})$.

This concludes the description of the algorithm. The analysis of $|H'|$ (the size analysis) is  analogous to the size analysis of the near-additive spanner from Section \ref{sec:sp_anal}. We omit it.
The size bound is $|H| = O_\kappa(n^{1+1/\kappa})$.
We next sketch the  analysis of its stretch and hopbound.

Consider a pair $u,v \in V$ of vertices such that $d_G(u,v) \in [R,2R]$, and let $\pi = \pi(u,v)$ be a shortest path between them. Observe that the sets $\{\cU_i \mid 0 \le i \le \ell-1\}$ form a partition of $V$, exactly as in the case of near-additive spanners.
We also write $\cU^{(i)} = \bigcup_{j \le i} \cU_j$, for all $i \in [0,\ell-1]$.

Next we provide upper bounds $R_i$ on the radii of clusters of $\cU_i$. Clusters of $\cU_0$ are singletons, and thus $R_0 = 0$. In general, for $i \in [0,\ell-2]$, we have $R_{i+1} = 3R_i + \delta_i$.
%(See Fig. ? for an illustration and intuition for this equation.)
Hence we have
$$R_{i+1} = \sum_{j=0}^i 3^j \delta_{i-j}= \delta_0/\eps^i \sum_{j=0}^i (3\eps)^j \le \delta_0/\eps^i {1 \over {1- 3\eps}}~.$$ For $\eps \le 1/6$, we have $R_{i+1} \le 2 \delta_0/\eps^i$.
In particular, $R_{\ell-1} \le 2\delta_0 (1/\eps)^{\ell-2} = 2R\eps$. (Recall that $R = \delta_0/\eps^{\ell-1}$.)

Moreover, it is easy to verify (by induction on $i$) that the radius of each cluster of $U_i$, for all $i \in [0,\ell-1]$, is attained by at most $i$ hops.

Let $z$ and $w$ be the leftmost and the rightmost $\cU_{\ell-1}$-clustered vertices on $\pi$, respectively.  Let $C_z$ and $C_w$ be the $\cU_{\ell-1}$-clusters of $z$ and $w$, respectively, and let $r_z$ and $r_w$ denote their respective centers. Observe that the hopset $H'$ contains a $z-w$ path obtained by going from $z$ to $r_z$ in $\ell-1$ hops or less, from $r_z$ to $r_w$ via direct edge of $H'$, and from $r_w$ to $w$ in at most $\ell-1$ additional hops.
The length of this path is at most
$$2R_{\ell-1} + d_G(r_z,r_w) \le 2R_{\ell-1} + d_G(z,w) + 2R_{\ell-1} = d_G(z,w) + 8\delta_0 \cdot (1/\eps)^{\ell-2}~.$$
Moreover, let $z'$ be the left neighbor of $z$ on $\pi$, and $w'$ be the right neighbor of $w$ on $\pi$. Then the path in $G \cup H'$ between $z'$ and $w'$, that starts with $G$-edge $(z',z)$, then takes the above hopset's $z-w$ path, and finally uses the $G$-edge $(w,w')$, has length at most $d_G(z',w') + 8\delta_0 \cdot (1/\eps)^{\ell-2}$, and uses at most $2 + 2(\ell-1) +1 = 2\ell+1$ hops. Hence the overhead of $8\delta_0 \cdot (1/\eps)^{\ell-2}$ can be charged
to the length of $\pi(u,w)$, which is at least $R = \delta_0 \cdot (1/\eps)^{\ell-1}$. This is a multiplicative overhead of $1+8\eps$.

Note also that the segments $u-z'$ and $w'-v$ of $\pi(u,v)$ contain vertices which are all clustered in $\cU^{(\ell-2)}$. (In other words, no vertex in these subpaths is $\cU_{\ell-1}$-clustered.) We divide these segments into subsegments of length at most $R \cdot \eps = \delta_0 \cdot (1/\eps)^{\ell-2} = \delta_{\ell-2}$.

In the case of hopsets this step requires more care than in the case of near-additive spanners, because in the latter case we dealt with {\em unweighted} graphs. Here the first segment starts at $u = x_0$, and ends in the last vertex $y = y_0$ along $\pi$ such that $d_G(x_0,y_0) \le R\eps$. If $d_G(x_0,y_0) < R\eps$, then the next segment starts in the right neighbor (with respect to $\pi$) $x_1$ of the vertex $y_0$. The edge $(y_0,x_1)$ is called a {\em connecting} edge between the two consecutive segments $x_0-y_0$ and $x_1-y_1$. Otherwise (if $d_G(x_0,y_0) = R\eps$), we set  $x_1 = y_0$. Then again $y_1$ is defined as the rightmost vertex with $d_G(x_1,y_1) \le R\eps$ along $\pi$, etc. This process continues until we reach $z'$.
The subpath between $w'$ and $v$ is divided into segments and connecting edges in the same manner.

In each such a segment $(x,y)$, we find the leftmost and the rightmost $\cU_{\ell-2}$-clustered vertices $z_{\ell-2}$ and $w_{\ell-2}$. Let $C_z$ and $C_w$ be their respective clusters, and $r_z$ and $r_w$ be their respective cluster centers.
Observe that the distance between $C_z$ and $C_w$ is at most $R\eps = \delta_{\ell-2}$, and thus their centers $r_z$ and $r_w$ were interconnected by a direct hopset edge $(r_z,r_w)$ in the interconnection step of phase $\ell-2$. The length of this hopset edge is
$$\omega(r_z,r_w) = d_G(r_z,r_w) \le d_G(r_z,z_{\ell-2}) + d_G(z_{\ell-2},w_{\ell-2}) + d_G(w_{\ell-2},r_w) \le
d_G(z_{\ell-2},w_{\ell-2}) + 2R_{\ell-2}~.$$
Hence there is a $z-w$ path  in the hopset that goes from $z$ to $r_z$, uses the edge $(r_z,r_w)$, and goes from $r_w$ to $w$.
Its length is at most
$$d_G(z_{\ell-2},w_{\ell-2}) + 4R_{\ell-2} \le d_G(z_{\ell-2},w_{\ell-2}) + 8 \delta_0 (1/\eps)^{\ell-3}~,$$
and it uses at most $2(\ell-2) + 1 = 2\ell-3$ hops.
Define $z'_{\ell-2}$ to be the left neighbor of $z_{\ell-2}$ on $\pi$, and $w'_{\ell-2}$ to be the right neighbor of $w_{\ell-2}$ on $\pi$. Then we also obtain a $z'_{\ell-2}-w'_{\ell-2}$ path in $G \cup H'$ with at most $2\ell-1$ hops and length at most
$d_G(z'_{\ell-2},w'_{\ell-2}) + 8\delta_0 (1/\eps)^{\ell-3}$.
We charge the overhead of $8\delta_0(1/\eps)^{\ell-3}$ to the length $\delta_0 \cdot (1/\eps)^{\ell-2} = R\eps$ of the segment between $x$ and $y$.\footnote{Strictly speaking, one needs also include the connecting edge incident  on $y$ in the segment.}
As a result the incurred stretch is at most $1+8\eps$ (on top of the stretch $1+8\eps$ incurred on the top-most, $(\ell-1)$st, level of the stretch analysis).

The number of hops incurred on the $(\ell-2)$nd level of the stretch analysis can be upper-bounded as follows.
There are $1/\eps$ segments, and on each of them $2\ell-1 = O(\ell)$ hops are used. (In addition, one hop per segment is used for connecting edges, but this is swallowed in the $O$-notation.) Hence the number of hops used by this level of stretch analysis is $O(\ell/\eps)$.

We then continue the stretch analysis in the same way, by zooming in into each of the subsegments $x-z'_{\ell-2}$ and $w'_{\ell-2}-y$. Their vertices are $\cU^{(\ell-3)}$-clustered, and thus on the next level of the stretch analysis we consider the leftmost and the rightmost $\cU_{\ell-3}$-clustered vertices on each subsegment of length at most $R \cdot \eps^2 = \delta_{\ell-3}$, etc.

The overall accumulated stretch on all the $\ell$ levels of the stretch analysis is thus $1+8\eps \ell$, and the overall number of hops can be crudely upper-bounded by $O(\ell) \cdot (1/\eps)^{\ell-1}$.
(To see this upper bound, observe that eventually we partition the path into $O((1/\eps)^{\ell-1})$ segments. The above analysis shows that on each segment at most  $O(\ell)$ hops are used.
To have a more precise bound, one notes that in fact on lower levels of the stretch analysis less hops per segment are used. This leads to a bound of $O((1/\eps)^{\ell-1})$.)

We also remark that no additive error is present here, even though the last segment may be shorter than $R\eps$.
This is because the entire path that we consider has length $\Theta(R)$, and thus the additive error of the last segment is swalllowed in the multiplicative stretch of $1+O(\eps)$. (This is unlike the case of near-additve spanners, where the original path may be very short, and so the additive error cannot be charged to the length of the original path.)

We now rescale $\eps' = O(\eps \ell)$, and obtain stretch of $1+ \eps'$ and hopbound $\beta = O\left({{\log \kappa} \over {\eps'}}\right)^{\log (\kappa+1) - 2}$. We summarize the result in the next theorem.

\begin{theorem} \cite{EN16}
\label{thm:hopsupinterconn}
For every pair of parameters $\eps > 0$ and $\kappa = 1,2,\ldots$, there exists $\beta = \beta(\eps,\kappa)
= O\left({{\log \kappa} \over \eps}\right)^{\log (\kappa +1)- 2}$, such that for every weighted undirected $n$-vertex
 graph $G = (V,E,\omega)$, there exists a $(1+\eps,\beta)$-hopset with $O_{\eps,\kappa}(n^{1+1/\kappa} \log \Lambda)$ edges.
\end{theorem}

As was remarked above, the $\log \Lambda$ factor in the size can be replaced by $\log n$ (see \cite{EN16}).

Note the striking similarity between Theorems \ref{thm:sp_supinterconn} and \ref{thm:hopsupinterconn}.

%%%%%%%%%%%%%%%%%%%%%%%%%%%%%
\section{Scale-Free Hopsets and Spanners}
\label{sec:scalefree}
%%%%%%%%%%%%%%%%%%%%%%%%%

In this section we present a universal extension of constructions from \cite{EP01,EN16}, described in Section \ref{sec:supinterconn}. They were developed in \cite{TZ06,EN19,HP17}.

%%%%%%%%%%%%%%%%%%%%%%%%%%%%%%%
\subsection{Construction}\label{sec:construction}
%%%%%%%%%%%%%%%%%%%%%%%%%%%%%%
Let $G = (V, E)$ be a graph with $n$ vertices (possibly with non-negative weights $w:E\to\R$ on the edges). Fix an integer parameter $\kappa \ge 1$ (it will be convenient to assume $\kappa=2^\ell-1$ for some integer $\ell\ge 1$). Denote $\ell=\log(\kappa+1)$.
Let $A_0,\dots, A_\ell$ be sets of vertices such that $A_0 = V$, $A_\ell = \emptyset$, and for $0 \leq i \leq \ell-2$, $A_{i+1}$ is created by sampling independently every vertex of $A_i$ with probability $q_i=n^{-2^i/\kappa}$.

For every $v\in V$ and $0\le i \le \ell-1$, define the pivot $p_i(v)$ as the closest vertex in $A_i$ to $v$, breaking ties in a consistent matter. For every $0\le i\le \ell-1$ and every $u \in A_i \setminus A_{i+1}$ define the bunch

$$\Bunch(v) = B(u) = \{v \in A_i ~:~ d_G(u, v) < d_G(u, A_{i+1})\} \cup \{p_{i+1}(u) \}~.$$

That is, the bunch $B(u)$ contains all the vertices which are in $A_i$ and closer to $u$ than $p_{i+1}(u)$, and  the level $i+1$ pivot.
We then define the emulator (and the hopset)
 $H = \{(u, v) ~:~ u \in V, v \in B(u)\}$, where the length $\omega'(u,v)$ of the edge $(u, v)$ is set as $d_G(u,v)$.

\paragraph{Size Analysis.} Fix any $0 \leq i \leq \ell-2$ and $u\in A_i\setminus A_{i+1}$, and consider the expected size of $B(u)$. If one orders the vertices of $A_i$ by their distance to $u$, then $B(u)$ contains the prefix of all the vertices in that ordering until the first one sampled to $A_{i+1}$. As this is a geometric random variable with parameter $q_i$, its expectation is $1/q_i=n^{2^i/\kappa}$. In addition, each vertex is connected to at most $\ell$ pivots, adding a term of $\ell n$.

Note that each $v\in V$ is included in $A_i$ with probability $\prod_{j=0}^{i-1}q_j$. These choices are independent for different vertices, so the expected size of $A_i$ is:

$$ N_i := \E[|A_i|] = n \prod_{j=0}^{i-1}q_j = n^{1-(2^i-1)/\kappa} ~.$$

So for each $0\le i\le\ell-2$, we have $N_i/q_i=n^{1+1/\kappa}$.
In addition, $N_{\ell-1}=n^{1-(2^{\ell-1}-1)/\kappa}=n^{(1+1/\kappa)/2}$, and it can be checked that
\[
\E[|A_{\ell-1}|^2]\le O(n^{1+1/\kappa})~.
\]

Note that for $u\in A_{\ell-1}$ we have $B(u)=A_{\ell-1}$, thus the expected size of the hopset $H$ is
\[
\sum_{i=0}^{\ell-2}N_i/q_i + E[|A_{k-1}|^2]  + \ell n =O(\log\kappa\cdot n^{1+1/\kappa})~.
\]

We remark that a more refined choice for the probabilities $q_i$ (and connecting to just 1 pivot, rather than all of them), can lead to size $O(n^{1+1/\kappa})$, essentially without affecting the other parameters.

%%%%%%%%%%%%%%%%%%%%%%%%%%%%%%%%%%%%
\subsection{Stretch Analysis of the Emulator}
\label{sec:emul_anal}
%%%%%%%%%%%%%%%%%%%%%%%%%%%%%%%%%%%%%%%

In this section we show that the edge set $H$ constructed above can serve as a universal near-additive emulator  for $G$.

Consider a pair of vertices $u,v \in V$. Let $\pi(u,v)$ be a shortest $u-v$ path. For some $\eps > 0$, we partition the path into segments of
length $(1/\eps)^{\ell-1}$, except the last segment that may be shorter. Each such a segment $x-y$ will be called a {\em level-$(\ell-1)$} segment. It will be further subdivided into level-$(\ell-2)$ segments of length $(1/\eps)^{\ell-2}$, etc. In general, for any $0 \le i \le \ell-1$,  level-$i$ segments have length $(1/\eps)^i$.\footnote{Except possibly one level-$i$ subsegment of  the possibly shorter level-$(\ell-1)$ segment; but this technicality has no real effect on the analysis.}

\begin{lemma}
\label{lm:emul_stretch}
There exist two universal constants $c,c' > 0$, such that for any $i$, $0 \le i \le \ell-1$, any $i$-level segment $x-y$ is either {\em successful},
i.e., satisfies
$$ (1) ~~d_H(x,y) \le d_G(x,y) + c \cdot i \cdot (1/\eps)^{i-1}~,$$
or {\em fails}, i.e., satisfies
$$ (2)~~ d_G(x,p_{i+1}(x)) \le c' \cdot (1/\eps)^i~.$$
\end{lemma}
\proof
The proof  is by induction on $i$.

{\bf Base:}  ($i = 0$)

Level $i =0$ segments have length 1, i.e., $(x,y) \in E$ is an edge. If $x \in A_1$, then $p_1(x) = x$, and so the segment fails ($d_G(x,p_1(x)) = 0$). Otherwise $x \in A_0 \setmns A_1$.

Then either $(x,y) \in H$, and then the segment is successful, as condition (1)  holds with 1 at the right-hand-side. Or, alternatively, $(x,y) \nin H$, i.e., $y \nin \Bunch(x)$. But then
$d_G(x,p_1(x)) \le d_G(x,y) =1$, proving condition (2) (i.e., the segment fails).

{\bf Step:}

Suppose that the assertion holds for all level-$i$ segments, for some $0 \le i \le \ell-2$. Consider a level-$(i+1)$ segment $x-y$. If all its level-$i$ subsegments are successful, then we concatenate the emulator's substitute paths for them. In the case that all its level-$i$ subsegments have length exactly $(1/\eps)^i$, the length of the  resulting path in the emulator  can be bounded by
\begin{eqnarray*}
d_H(x,y) & \le & 1/\eps \cdot ((1/\eps)^i + c \cdot i \cdot (1/\eps)^{i-1}) \\
& = & (1/\eps)^{i+1} + c \cdot i \cdot (1/\eps)^i~.
\end{eqnarray*}
In the general case, one obtains here an upper bound of $d_G(x,y) + c \cdot i \cdot (1/\eps)^i$, by essentially the same argument. Hence in this case the segment $u-v$ is successful as well.

Otherwise there are some failing level-$i$ subsegments of $x-y$. Let $x_L-y_L$ and $x_R-y_R$ be the leftmost and the rightmost such subsegments.   Let $z_L = p_{i+1}(x_L)$, $z_R = p_{i+1}(y)$. By the inductive hypothesis, we have
$d_G(x_L,z_L), d_G(y_R,z_R) \le c' \cdot (1/\eps)^i$.
Then
\begin{eqnarray*}
d_G(z_L,z_R) &\le& d_G(z_L,x_L) + d_G(x_L,y_R) + d_G(y_R,z_R) \\
& \le & d_G(x_L,y_R) + 2c' \cdot (1/\eps)^i~.
\end{eqnarray*}
Observe that the edges $(x_L,z_L)$, $(y_R,z_R)$ belong to the emulator $H$. If $(z_L,z_R) \in H$ as well, then
\begin{eqnarray*}
d_H(x_L,y_R)  & \le & d_H(x_L,z_L) + d_H(z_L,z_R) + d_H(z_R,y_R) \\
& = & d_G(x_L,z_L) + d_G(z_L,z_R) + d_G(z_R,y_R) \\
& \le & d_G(x_L,y_R) + 2(d_G(x_L,z_L) + d_G(z_R,y_R)) ~\le~ d_G(x_L,y_R) + 4 c' \cdot (1/\eps)^i~.
\end{eqnarray*}
Also, note that each of the level-$i$ segments of the subpaths $x-x_L$ and $y_R-y$ of the segment $x-y$ are
successful, and there are
$${{d_G(x,x_L) + d_G(y_R,y)} \over {(1/\eps)^i}}$$ such segments. Hence
\begin{eqnarray*}
d_H(x,x_L) + d_H(y_R,y) & \le & {{d_G(x,x_L) + d_G(y_R,y)} \over {(1/\eps)^i}} \cdot ((1/\eps)^i + c \cdot i \cdot (1/\eps)^{i-1}) \\
& = & (d_G(x,x_L) + d_G(y_R,y)) \cdot (1 + c \cdot i \cdot \eps)~.
\end{eqnarray*}
Thus we have
\begin{eqnarray*}
d_H(x,y) & \le & d_H(x,x_L) + d_H(x_L,y_L) + d_H(y_R,y) \\
& \le & (d_G(x,x_L) + d_G(y_R,y))\cdot ( 1 + c \cdot i \cdot \eps) + d_G(x_L,y_R) + 4c' \cdot (1/\eps)^i \\
& \le & d_G(x,y) (1 + c \cdot i \cdot \eps) + 4c' \cdot (1/\eps)^i~.
\end{eqnarray*}
Observe that $d_G(x,y) = (1/\eps)^{i+1}$, i.e., $4c' (1/\eps)^i = 4c' \eps \cdot d_G(u,v)$.
Hence in this case
$$d_H(x,y) \le (1 + c \cdot i \cdot \eps + 4c' \cdot \eps) \cdot d_G(x,y)~.$$
For $c = 4c'$, we obtain that
$$d_H(x,y) \le (1 + c(i+1)\eps)(1/\eps)^{i+1}~,$$ and thus the segment $x-y$ is successful.

Otherwise $(z_L,z_R) \nin H$, i.e., $z_R \nin \Bunch(z_L)$.
Then we have
$$d_G(z_L,p_{i+1}(z_L)) \le d_G(z_L,z_R) \le d_G(x_L,y_R) + 2 c' (1/\eps)^i~.$$
Hence
\begin{eqnarray*}
d_G(x,p_{i+1}(x)) & \le & d_G(x,p_{i+1}(z_L)) \le d_G(x,x_L) + d_G(x_L,z_L) + d_G(z_L,p_{i+1}(z_L)) \\
& \le &
d_G(x,x_L) + c' \cdot (1/\eps)^i  + d_G(x_L,y_R) + 2c' (1/\eps)^i \\
& \le & d_G(x,y) + 3c'(1/\eps)^i = (1/\eps)^{i+1} + 3c' (1/\eps)^i ~\le~ c'(1/\eps)^{i+1}~.
\end{eqnarray*}
The last inequality holds for $c' \ge {1 \over {1 - 3\eps}}$. Hence if we set  $c' \ge 2$, it holds for all $\eps < 1/6$.

This completes the proof.
\QED

Observe that an $(\ell-1)$-level segment $x-y$ cannot fail, and thus we have
$d_H(x,y) \le d_G(x,y)(1 + c (\ell-1) \cdot \eps$.
By concatenating the emulator's substitute paths for all the segments, we obtain  that for any $u,v \in V$,
$$d_H(u,v) \le (1 +c (\ell-1) \cdot \eps) d_G(u,v) + O(c (\ell-1) (1/\eps)^{\ell-1})~.$$
(Exactly as in Section \ref{sec:supinterconn}, the additive term is because of the last segment.)

By rescaling $\eps' = c(\ell-1)\eps$, we obtain that the stretch of the emulator is $(1+\eps,O\left({{\log \kappa} \over \eps}\right)^{\log (\kappa+1) - 2})$. Note also that this construction does not accept $\eps$ as a parameter, and thus applies to all $\eps > 0$.

We summarize this analysis with the following theorem due to \cite{TZ06}. (The proof that we provided is however different from the original proof from \cite{TZ06}.)

\begin{theorem} \cite{TZ06}
For any $\kappa = 1,2,\ldots$, and any $n$-vertex graph $G = (V,E)$, the graph $G'=(V,H,\omega')$ constructed as above is a $(1+\eps,\beta(\eps,\kappa))$-emulator for $G$ with $O_\kappa(n^{1+1/\kappa})$ edges, for all $\eps < 1/6$, where $\beta = \beta_{EP}$.
\end{theorem}

%%%%%%%%%%%%%%%%%%%%%%%%%%%%%
\subsection{Stretch Analysis of the Hopset}
\label{sec:scalefree_hopset}
%%%%%%%%%%%%%%%%%%%%%%%%%%%%%%

In this section we show that the very same edge set $H$ constructed in the beginning of this section provides a $(1+\eps,\beta)$-hopset (naturally, of the same size), even for weighted graphs.

Again, consider a shortest $u-v$ path $\pi(u,v)$. Denote $L = \omega(\pi(u,v))$. We partition it into $1/\eps$ segments of length $L\eps$ each. (Suppose for simplicity that it can be divided into segments of precisely this length. If it is not the case, it can be taken care of, essentially without affecting the analysis.) Those segments are again subdivided to $1/\eps$ subsegments of length $L \cdot \eps^2$ each, etc, for $\ell-1$ levels.
Segments of length $(L \cdot \eps^{\ell-1}) \cdot (1/\eps)^i$ are the $i$-level segments.
So, in a sense, $\gamma = L \cdot \eps^{\ell-1}$ is the ``distance unit" of the construction. See also Section \ref{sec:supinterconn}.

One can assume that all weights are greater or equal to 1. Assume also that $L \ge (1/\eps)^{\ell-1}$.
If it is not the case, the graph $G$ itself has a $u-v$ path of length $d_G(u,v)$ with at most $(1/\eps)^{\ell-1}$ hops.

The next lemma is completely analogous to Lemma \ref{lm:emul_stretch}.

\begin{lemma}
\label{lm:hopset_stretch}
There exist universal constants $c,c' > 0$ such that for any $0 \le i \le \ell-1$, any $i$-level segment $x-y$ is either {\em successful}, i.e., satisfies
$$(1)~~ d_H^{(1/\eps)^i)}(x,y) \le \gamma \cdot \left((1/\eps)^i + c i \cdot (1/\eps)^{i-1}\right)~,$$
or {\em fails}, i.e., satisfies
$$ (2)~~ d_G(x,p_{i+1}(x)) \le \gamma \cdot c' \cdot (1/\eps)^i~.$$
\end{lemma}
\proof
The proof is again  by induction on $i$.

{\bf Base}: ($i = 0$)  \\
If $x \in A_1$, then the segment satisfies (2) with 0 in the right-hand-side. Otherwise $x \in A_0 \setmns A_1$. If $(x,y) \in H$ then the segment is successful. Otherwise, $d_G(x,p_1(x)) \le d_G(x,y) = \gamma$, and the segment fails. In both cases the assertion of the lemma holds.

\begin{figure}
	\begin{center}
		\includegraphics[width=1\textwidth]{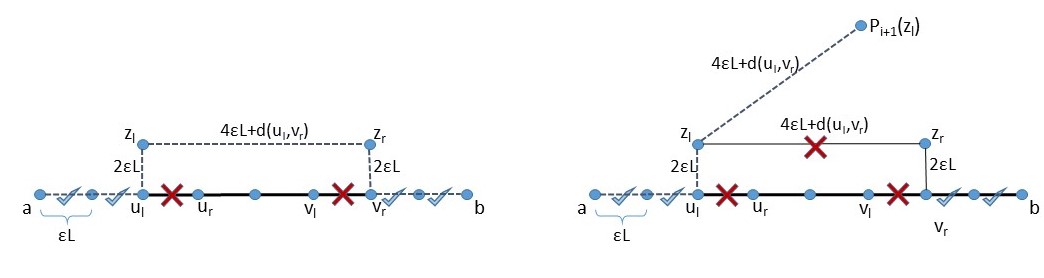}
		\caption{\small An illustration for the $1+\epsilon$ stretch version. Above are the two cases when considering an interval $[a,b]$ of length $L$ at level $i$, which is divided to $1/\epsilon$ sub-intervals (the case when all sub-intervals are successful is omitted). The dashed line represents the path in $G\cup H$ we find.  On the left is the case that some sub-intervals failed, and there is an hopset edge between the level $i$ pivots of the leftmost and rightmost failed intervals' endpoints; in this case we have a $1/\epsilon^i$-hops path with stretch $1+ci\epsilon$. The other case is that there is no such edge, but then we see a level $i+1$ pivot at distance at most $c'L$.}
		\label{fig:hopset-eps}
	\end{center}
\end{figure}

{\bf Step:} The proof of the induction step is completely analogous to the proof of the induction step of Lemma \ref{lm:emul_stretch}, except that all expressions need to be scaled up by a factor of $\gamma$.
An illustration is provided in Figure \ref{fig:hopset-eps}.

\QED

Lemma \ref{lm:hopset_stretch} implies the following theorem.

\begin{theorem} \cite{EN19,HP17}
For any $\kappa = 1,2\ldots$, and any $n$-vertex weighted graph $G = (V,E,\omega)$, the graph $G' = (V,H,\omega')$ constructed above is a $(1 + \eps,\beta)$-hopset for $G$ with $O_\kappa(n^{1+1/\kappa})$ edges,
and $\beta = \beta_{EP}$.
\end{theorem}

%%%%%%%%%%%%%%%%%%%%%%%%%%%%%%%%%%%%
\section{Conclusions and Open Problems}
\label{sec:concl}
%%%%%%%%%%%%%%%%%%%%%%%%%%%%%%%%%%%%%%%

As we have seen, there is a striking similarity not just between the results concerning near-additive spanners for unweighted graphs and near-exact hopsets for weighted ones, but also between the techniques used to construct them and to analyze these constructions.
Specifically, the superclustering and interconnection approach (see Section \ref{sec:supinterconn}) due to \cite{EP01} gives rise to very similar constructions of these two objects \cite{EP01,EN16}, and this is also the case with its scale-free extension due to \cite{TZ06} (see \cite{EN19,HP17} and Section \ref{sec:scalefree}).

The situation is similar in the case of Cohen's approach \cite{C94} that relies on pairwise covers \cite{C93,ABCP93}. This approach also gives rise to closely related constructions and analyses for both near-exact hopsets \cite{C94} and near-additive spanners \cite{E01,EZ04}. This approach was left out of the scope of the current survey.

A very interesting open problem is to explain the relationship between near-additive spanners and near-exact hopsets rigorously, i.e., by providing a reduction between these two objects.

Another major open question is to determine the correct dependency of $\beta$ on $\eps$ and $\kappa$ for both spanners and hopsets. Can one achieve $\beta$ polynomial in $\kappa$ for near-additive spanners and/or near-exact hopsets?

Numerous related open problems arise if one allows a larger stretch than $1+\eps$. Currently there are known constructions with stretch $3+\eps$ and $\beta$ polynomial in $\kappa$ \cite{Pet07,EGN19,BLP19}. Can this be achieved with stretch smaller than 3? What is the right three-way tradeoff between the sparsity parameter $\kappa$, the multiplicative stretch $\alpha$ and the hop-bound (or additive stretch) $\beta$?

We have also pointed out that the current state-of-the-art constructions of {\em universal} near-additive spanners  (see Section \ref{sec:scalefree}) lag behind their non-universal counterparts. Specifically, the state-of-the-art bound on the parameter $\beta$ in the universal constructions \cite{EN17,Pet07} is $\beta_{EP}^{\log_{4/3} 2}$, where
$\beta_{EP} = \left({{\log \kappa} \over \eps}\right)^{\log \kappa-2}$ is the state-of-the-art bound for non-universal constructions \cite{EP01}. Narrowing this gap, or proving a lower bound precluding this, is an open problem.

In this survey we focused on {\em existential}, i.e., combinatorial properties of near-additive spanners and near-exact hopsets. However, for many applications it is important to compute them efficiently in various computational models. For example, in the centralized model of computation one introduces a control parameter $\rho > 0$, and can obtain
$(1+\eps,\beta)$-spanners with $O_{\eps,\kappa}(n^{1+1/\kappa})$ edges and
$$\beta = \left({{\log \kappa \rho + 1/\rho} \over \eps}\right)^{\log\kappa \rho +1/\rho}$$ in time
$O(|E| \cdot n^\rho)$ \cite{EN16,EN17,E01,EZ04}. The tradeoff looks similarly in other models of computation, i.e., the overhead of $n^\rho$ in the running time  at the expense of larger $\beta$ is persistent.  Improving upon this tradeoff is an open problem. Its positive resolution is likely to lead to improved algorithms for the computation of approximate shortest paths, distributed routing tables, parallel distance oracles, and other applications.

Finally, in many applications of hopsets one needs not just approximate distances, but also paths that implement these distances. For this aim, {\em path-reporting} hopsets, i.e., hopsets from which approximate paths can be readily retrieved were introduced in \cite{EN16}. Their parameters are, however, somewhat inferior to those of their non-path-reporting counterparts. Devising path-reporting hopsets with improved parameters is also an interesting open problem.

%%%%%%%%%%%%%%%%%%%%%%%%%%%%%%%%%%%%%%%%%%%%%%%%%%%%%%%%%%%%%%%%%%%%%%%%%%
\bibliographystyle{alpha}
\bibliography{hopset}
%%%%%%%%%%%%%%%%%%%%%%%%%%%%%%%%%%%%%%%%%%%%%%%%%%%%%%%%%%%%%%%%%%%%%%%%%%

\end{document}